\documentclass[preprint]{revtex4-2}

\usepackage{graphicx}
\usepackage{amssymb}
\usepackage{amsmath}
\usepackage{color,xcolor}
\usepackage{bm}
\usepackage{physics}
\usepackage{ulem}

\usepackage{changes}

\begin{document}


\title{Nonreciprocity induced spatiotemporal chaos: Reactive vs dissipative routes}

\author{Jung-Wan Ryu}
    \address{Center for Theoretical Physics of Complex Systems, Institute for Basic Science (IBS), Daejeon 34126, Republic of Korea}
    \address{Basic Science Program, Korea University of Science and Technology (UST), Daejeon 34113, Republic of Korea}
\date{\today}

\begin{abstract}
Nonreciprocal interactions fundamentally alter the collective dynamics of nonlinear oscillator networks. Here we investigate Stuart–Landau oscillators on a ring with nonreciprocal reactive or dissipative couplings combined with Kerr-type or dissipative nonlinearities. Through numerical simulations and linear analysis, we uncover two distinct and universal pathways by which enhanced nonreciprocity drives spatiotemporal chaos. Nonreciprocal reactive coupling with Kerr-type nonlinearity amplifies instabilities through growth-rate variations, while nonreciprocal dissipative coupling with Kerr-type nonlinearity broadens eigenfrequency distributions and destroys coherence, which, upon nonlinear saturation, evolve into fully developed chaos. In contrast, dissipative nonlinearities universally suppress chaos, enforcing bounded periodic states. Our findings establish a minimal yet general framework that goes beyond case-specific models and demonstrate that nonreciprocity provides a universal organizing principle for the onset and control of spatiotemporal chaos in oscillator networks and related complex systems.
\end{abstract}

\maketitle

\section{Introduction}

Nonreciprocal coupling, also referred to as asymmetric or directed interactions, 
has emerged as a unifying principle across physics, biology, ecology, and photonics. 
The concept of an ``order of life,'' where asymmetry organizes collective behaviors, 
has long been discussed in ecological contexts. Eco--evolutionary theory has shown that 
nonreciprocal interactions sustain perpetual Red Queen dynamics and drive endless 
evolutionary change in multispecies communities~\cite{VanValen1973, Dieckmann1999Nature, Morran2011Science}. 
This perspective has been formalized in consumer--resource models and metaecosystems~\cite{Tadiri2024FE}, 
and extended to vegetation fronts destabilized by nonreciprocity~\cite{PintoRamos2025}. 
More recent studies demonstrate that asymmetry not only sustains continuous arms races 
but also enables genuinely open--ended Red Queen dynamics, where novel adaptive states 
emerge without bound~\cite{Mahadevan2025PRXLife}, while a universal niche geometry 
governs ecosystem responses to perturbations~\cite{Goyal2025PRXLife}. 
This concept has further been extended to physical systems through studies of 
nonreciprocal interactions~\cite{Mandal_Sen_Astumian_Chem_2024, Dinelli_NatComm_2023, Zhang_PRResearch_2023}. 
At the microscopic level, active matter provides archetypal realizations where 
nonreciprocity breaks detailed balance. Scalar active mixtures described by 
nonreciprocal Cahn--Hilliard models exhibit traveling bands, oscillatory instabilities, 
and self--propelled patterns~\cite{Saha2020PRX, You2020PNAS, Saha2025Effervescence}, while 
field--theoretic and related approaches reveal nonreciprocal pattern formation of 
conserved fields~\cite{Brauns2024PRX, Fruchart2021Nature} and heterogeneous 
self--organization~\cite{CarlettiMuolo2022CSF164_112638}. 
Together, these studies suggest that nonreciprocity provides a common framework 
linking biological evolution, ecological interactions, and physical active matter systems.

In physical systems, nonreciprocity underlies diverse dynamical phenomena. 
Asymmetric couplings drive front propagation in optical chains, modulational 
instabilities in nonlinear Hatano--Nelson lattices, and nonreciprocal entanglement 
in superconducting circuits~\cite{AguileraRojas2024CP, Longhi2025APR, Ren2024arxiv}. 
They also generate chaos in resonators, optomechanical platforms, and Duffing 
oscillator chains~\cite{Zhang2021_PRA_SpinningResonator, Zhang2025SqueezingChaos, 
Lai2025ChaoticDiodeDuffing}. Photonic and non-Hermitian lattices further realize 
nonreciprocity, from modulation-induced Hatano--Nelson couplings to spatially offset 
excited states~\cite{Orsel2025PRL134_153801, Jiang2025PRA111_052218}. Earlier studies 
showed that asymmetric coupling can fundamentally alter synchronization and chaos in 
extended systems~\cite{Bragard2003PRL91_064103, Nishikawa2006PRE}, while more recent work 
demonstrated robust non-Hermitian and even quantum synchronization under directed 
interactions~\cite{Zhang2025AdvSci, Kehrer2025QuantumSyncBlockade, Ho2024PNAS}. 
These studies collectively establish that nonreciprocity is not a marginal correction 
but a fundamental mechanism for organizing collective dynamics across disciplines, 
and its impact has been observed from classical to quantum regimes~\cite{PRX2025_QuantumSpins}.

In oscillator networks, nonreciprocal coupling decisively shapes collective dynamics. 
Unidirectional interactions can induce global amplitude death and stabilize homogeneous 
steady states~\cite{Ryu2017_Chaos_AD}, while synchronization theory shows that maximally synchronizable networks are inherently directed~\cite{Nishikawa2006PRE}. In both phase-reduced and Stuart--Landau models, asymmetric coupling gives rise to multistability, chimera patterns, and explosive oscillation death~\cite{Manoranjani2021_Chaos, Jaros2021_Chaos, Mendola2025_PRE}. These results establish oscillator models as a minimal yet versatile framework where nonreciprocity organizes complex states and spatiotemporal chaos. Building on this foundation, our work identifies two universal routes through which nonreciprocity drives chaotic dynamics in oscillator networks.

Here we focus on how strengthening nonreciprocity organizes spatiotemporal chaos 
in oscillator networks. We show that chaos arises via two universal routes: 
instability amplification through nonreciprocal reactive couplings with Kerr-type nonlinearity, and coherence loss through nonreciprocal dissipative couplings with Kerr-type nonlinearity. In both cases, nonlinear saturation transforms linear instabilities or incoherence into fully developed chaos, whereas dissipative nonlinearities suppress chaotic growth and yield only bounded oscillations. This systematic comparison highlights that nonreciprocity is not a minor perturbation but a key organizer of chaotic dynamics.

The paper is organized as follows. Sec.~II introduces the model and methods. 
Sec.~III describes dynamical regimes in four representative cases (reactive and 
dissipative couplings combined with Kerr-type or dissipative nonlinearities). 
Sec.~IV concludes with a summary and outlook.

\section{Model and Methods}

The Stuart--Landau equation, which describes the behavior of a nonlinear oscillator near the Hopf bifurcation, has been widely used to explain the dynamical behaviors of chemical, biological, classical, and quantum oscillator systems. The well-known equation is expressed as 
\begin{equation}
\dot{z} = (\mu + i \omega - \xi |z|^2) z .
\end{equation}
Here $z$ is the complex amplitude, $\mu$ controls the onset of oscillations, and $\omega$ is the intrinsic angular frequency of an oscillator. $\xi$ is the complex coefficient of the cubic nonlinearity: its real part corresponds to \textit{dissipative nonlinearity}, which provides nonlinear damping for $\mathrm{Re}(\xi) > 0$ (amplitude saturation) or nonlinear anti-damping for $\mathrm{Re}(\xi) < 0$, while its imaginary part corresponds to \textit{Kerr-type nonlinearity}, which induces nonlinear frequency shifts. This distinction forms the basis for the four representative cases analyzed in Sec.~III.

We consider a one-dimensional ring network of $N$ coupled identical Stuart--Landau oscillators, which serves as a minimal model to capture the interplay between nonlinearity and nonreciprocity. Each oscillator is described by a complex amplitude $z_j(t)$, where $j=1,\dots,N$, and evolves according to
\begin{equation}
\dot{z}_j = (\mu + i \omega - \xi |z_j|^2) z_j 
+ K \left( J_L z_{j+1} + J_R z_{j-1} \right),
\label{eq:SLO}
\end{equation}
with periodic boundary conditions $z_{j+N} = z_j$. $K$ is the overall coupling constant and the coefficients $J_L$ and $J_R$ control the strength of coupling to the left and right neighbors, respectively. Without coupling ($K = 0$), $N$ oscillators that have stable fixed points at the origin approach the origin asymptotically after a transient for negative $\mu$. The instability does not appear without coupling. In the following, we set $\mu = -0.5$, $\omega = 2$, $K = 1$, $J_L + J_R = 1$, and $N = 100$ unless otherwise stated.

Reciprocal coupling corresponds to the symmetric case $J_L = J_R$, 
whereas nonreciprocal coupling arises when $J_L \neq J_R$. 
Throughout this work, we use the terms \textit{nonreciprocity} and \textit{asymmetric coupling} interchangeably, 
emphasizing that the broken symmetry between left and right interactions renders the network nonreciprocal. 
Two types of nonreciprocity are distinguished: 
(i) \textit{reactive coupling}, when $K$ is purely imaginary, leading to shifts of oscillation frequencies; and 
(ii) \textit{dissipative coupling}, when $K$ is real, leading to amplification or damping of oscillation amplitudes. 

To characterize the system we analyze three complementary quantities: 
(i) spatiotemporal patterns $|z_j(t)|$ and their Euclidean norms over time; 
(ii) time series of representative oscillators to probe local dynamics; and 
(iii) eigenvalue spectra of the Jacobian matrix linearized about the fixed point, 
which explain the transition from stability to instability or from coherence to incoherence and the subsequent nonlinear regimes. This combined approach allows us to connect microscopic coupling asymmetries to macroscopic chaotic patterns in a systematic way.

\section{Results}

Having established the model and analysis framework, 
we now present the dynamical behavior of the Stuart--Landau network 
under different combinations of nonreciprocal coupling and nonlinearity. 
Our focus is on four representative cases: 
(i) reactive coupling with Kerr-type nonlinearity, 
(ii) dissipative coupling with Kerr-type nonlinearity, 
(iii) reactive coupling with dissipative nonlinearity, and 
(iv) dissipative coupling with dissipative nonlinearity. 
In each case, we examine spatiotemporal patterns and corresponding Euclidean norms, 
time series of representative oscillators, and Jacobian eigenvalue spectra to elucidate the underlying mechanisms. This systematic comparison reveals two universal routes by which nonreciprocity induces spatiotemporal chaos.

\begin{figure*}[!t]
    \centering
    \includegraphics[width=1.0\linewidth]{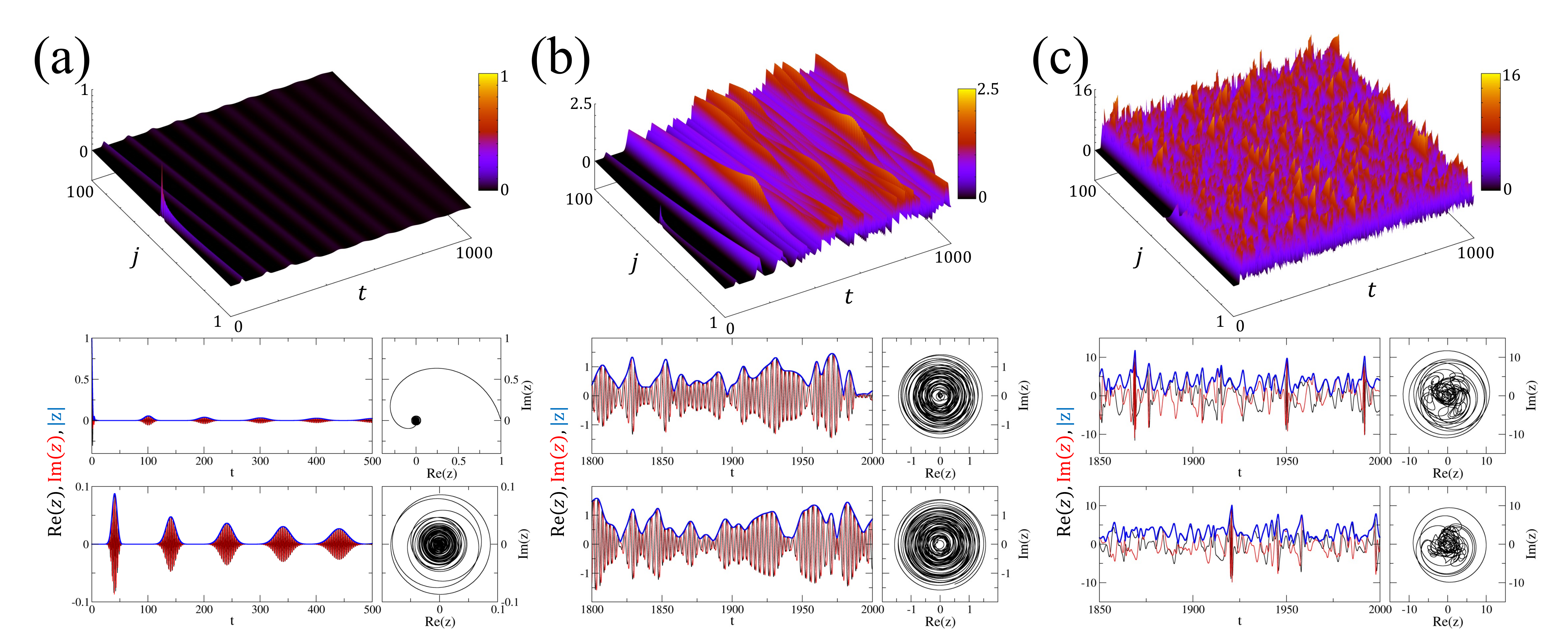}
    \caption{Reactive coupling with Kerr-type nonlinearity. Spatiotemporal patterns of the amplitudes, time series of individual oscillators, and trajectories in the complex plane for a ring of 100 Stuart--Landau oscillators. Parameters are $\mu=-0.5$, $\omega=2.0$, $\xi=0.1i$, and $K=1.0i$. Panels show (a) $J_L=0.75$, (b) $J_L=0.76$, and (c) $J_L=1.0$. In panel (a), the time series and trajectories are plotted from the initial condition, while in (b) and (c) the initial transient has been removed. The diagonal stripes in the amplitude space--time plots indicate that the underlying phase wavefront propagates around the ring at a constant velocity, corresponding to a traveling-wave (rotating-wave) state in (a) and (b).}
    \label{fig:fig1}
\end{figure*}

\begin{figure}[!t]
    \centering
    \includegraphics[width=0.5\linewidth]{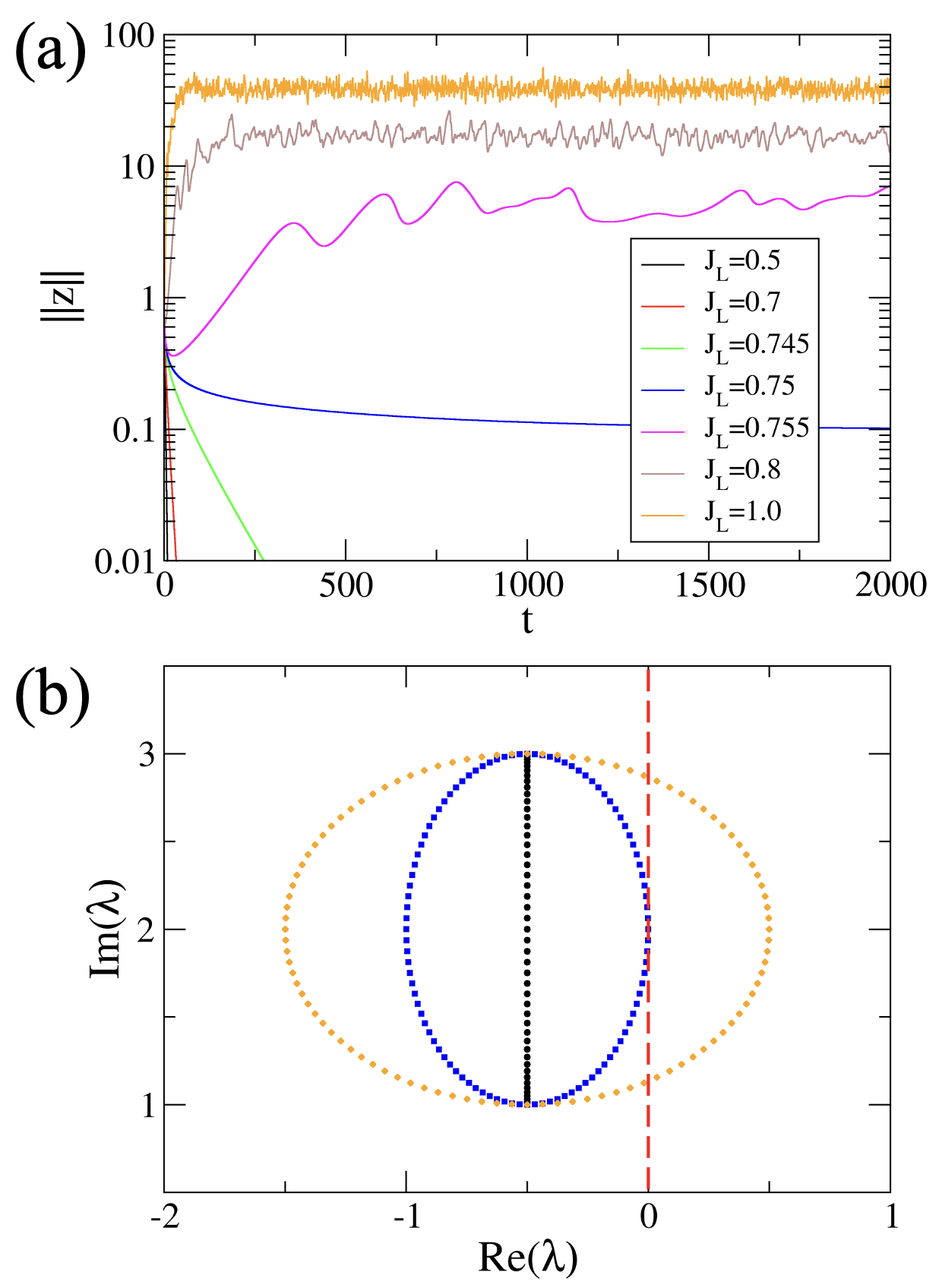}
    \caption{(a) Time evolution of the Euclidean norm and (b) complex eigenvalues of Jacobian matrix. The time evolution of the global Euclidean norm, Eq.~(\ref{eq:EucNorm}), corresponding to the spatiotemporal patterns in Fig.~\ref{fig:fig1}. The curves illustrate a distinct pathway to spatiotemporal chaos in the reactive--Kerr case. There is transition between decaying behaviors and bounded fluctuation dynamics corresponding to convective decaying states and spatiotemporal chaos, respectively. Parameters are the same as in Fig.~\ref{fig:fig1}.}
\label{fig:fig2}
\end{figure}

\subsection{Reactive coupling with Kerr-type nonlinearity}

We first consider the case of reactive coupling with Kerr-type nonlinearity, where the coupling constant is purely imaginary ($K = 1.0 i$), corresponding to reactive coupling, and the nonlinear coefficient is purely imaginary ($\xi = 0.1 i$). The reactive coupling perturbs frequencies of uncoupled oscillators. The coupling strength $|K|$ widens the distributions of frequencies of oscillators in the coupled network. This setting demonstrates how convective instabilities, triggered by coupling asymmetry, exhibit the transition to spatiotemporal chaos under the influence of nonlinearity. 

Figure~\ref{fig:fig1} summarizes the dynamics. 
The initial condition is $z_j = 1.0$ for $j=50$ and $z_j = 0.0$ otherwise. 
For the symmetric case ($J_L = J_R = 0.5$), the system exhibits global amplitude death due to the stable fixed point by negative $\mu$. The symmetric coupling does not make system unstable. With small asymmetry, the spatiotemporal pattern displays convective decay, defined as the directional decrease of oscillation amplitude during propagation. As $J_L$ increases beyond $0.75$, convective amplification emerges, leading to irregular traveling (rotating) waves that mark the onset of convective instability. When the amplitude grows beyond the threshold, nonlinear frequency shifts induced by the Kerr term suppress further amplification and convert the instability into 
spatiotemporal chaos. 
Examining the time traces of representative oscillators (j = 50 and j = 10) reveals irregular bounded oscillations, indicating the loss of traveling waves.
Larger asymmetry yields more complicated time evolution and wider amplitude excursions. 

We introduce the Euclidean norm as a global measure of network activity,
\begin{equation}
||Z(t)|| = \sqrt{\sum_{j=1}^{N} |z_j(t)|^2} .
\label{eq:EucNorm}
\end{equation}
This quantity captures the overall growth and suppression of oscillations as well as fluctuation of the time evolutions originated from spatiotemporal chaos, complementing the spatiotemporal patterns and Jacobian spectra.

Figure~\ref{fig:fig2} (a) shows the global dynamics corresponding to the spatiotemporal patterns in Fig.~\ref{fig:fig1}. The norm decreases exponentially due to convective decay when $J_L < 0.75$ but does algebraically near the critical value $J_L \lesssim 0.75$, reflecting critical slowing down. When $J_L > 0.75$, the norm initially grows rapidly due to convective amplification (also known as directional amplification) \cite{Wanjura_NatComm_2020,Ryu2023PRA_Dynamics}, but then saturates at an irregular level once modulational instability sets in, producing saturation-induced suppression of divergence. 

To interpret the onset of instability and the decoherence of oscillators, we analyze the eigenvalues of the Jacobian matrix at the origin. The real parts of the eigenvalues correspond to decay or growth rates, and the imaginary parts correspond to the angular frequencies of oscillators near the origin. Instability occurs once the maximal real part becomes positive, and the distribution of imaginary parts is related to the eigenfrequency coherence of oscillators. The Jacobian matrix at the origin for Eq.~(\ref{eq:SLO}) is
\begin{equation}
\label{eq:Jacobian}
J = \left(\begin{array}{ccccc}
 \mu + i \omega & K J_L & \cdots & 0 & K J_R \\
 K J_R & \mu + i \omega & \cdots & 0 & 0 \\
 \vdots & \vdots & \ddots & \vdots & 0 \\
 0 & 0 & \cdots & \mu + i \omega & K J_L \\
 K J_L & 0 & \cdots & K J_R & \mu + i \omega 
\end{array}\right),
\end{equation}
which is an $N\times N$ circulant matrix with first row $(\mu+i\omega,\;KJ_L,\;0,\dots,0,\;KJ_R)$.
For Fourier wavenumbers $q_k=\tfrac{2\pi k}{N}$ $(k=0,\dots,N-1)$, the eigenvalues are
\begin{equation}
\lambda_k=\mu+i\omega
+K\!\left(J_L e^{+iq_k}+J_R e^{-iq_k}\right).
\end{equation}
The associated eigenvectors are Fourier modes of the form $\exp(\pm i q_k j)$. When $J_L \neq J_R$, the imbalance between the two directions selects a net propagation, corresponding to a traveling-wave (rotating-wave) state circulating around the ring. In contrast, for the symmetric case $J_L = J_R$, the $\pm q_k$ components combine with equal weight to form a standing wave $\cos(q_k j)$.
Since $J_L+J_R=1$, this may be rewritten as
\begin{equation}
\lambda_k=\mu+i\omega+K\!\left(\cos q_k+i\Delta J\,\sin q_k\right),
\qquad \Delta J:=J_L-J_R.
\end{equation}
When the coupling constant is purely imaginary, $K = i b$ with $b \in \mathbb{R}$, corresponding to reactive coupling, the eigenvalues take the form
\begin{align}
\mathrm{Re}(\lambda_k) &= \mu - b\,\Delta J \sin q_k, \tag{7} \\
\mathrm{Im}(\lambda_k) &= \omega + b \cos q_k. \tag{8}
\end{align}
Here the imaginary parts vary within the fixed interval $[\omega-|b|,\omega+|b|]$. 
In the symmetric case $\Delta J = 0$, all eigenvalues align on a vertical line at $\mathrm{Re}(\lambda_k) = \mu$, so that the frequencies disperse while the growth rate remains fixed [Fig.~\ref{fig:fig2}(b)]. 
With asymmetry, however, the real parts become $q_k$-dependent, so that the vertical line deforms into an ellipse elongated along the real axis. 
As $\Delta J$ increases, the real parts broaden their spread in proportion to $\Delta J$, whereas the imaginary parts preserve their variation within the constant interval $[\omega-|b|,\omega+|b|]$.

The Jacobian eigenvalue spectrum provides a clear explanation for the transition to spatiotemporal chaos by nonreciprocity. Once the maximal real part crosses zero, linear instability arises. At this point, Kerr-type nonlinearity prevents unbounded divergence originated from nonreciprocal reactive coupling and produces fully developed spatiotemporal chaos.

\begin{figure*}[!t]
    \centering
    \includegraphics[width=1.0\linewidth]{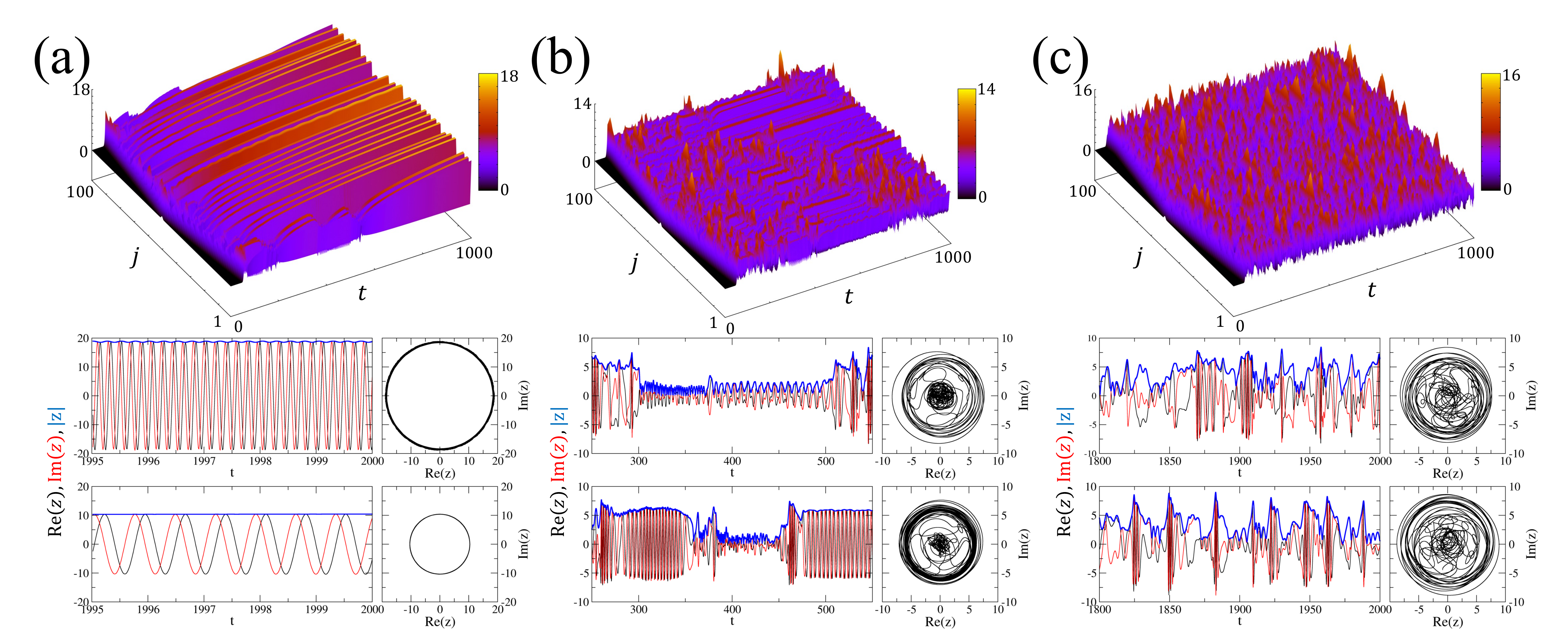}
    \caption{Spatiotemporal patterns of the amplitudes, time series of individual oscillators, and trajectories in the complex plane for a ring of 100 Stuart--Landau oscillators. Parameters are $\mu = -0.5$, $\omega = 2.0$, $\xi = 0.1 i$, and $K = 1.0$. Panels show (a) $J_L = 0.50$, (b) $J_L = 0.515$, and (c) $J_L = 0.55$.}
    \label{fig:fig3}
\end{figure*}

\begin{figure}[!t]
    \centering
    \includegraphics[width=0.5\linewidth]{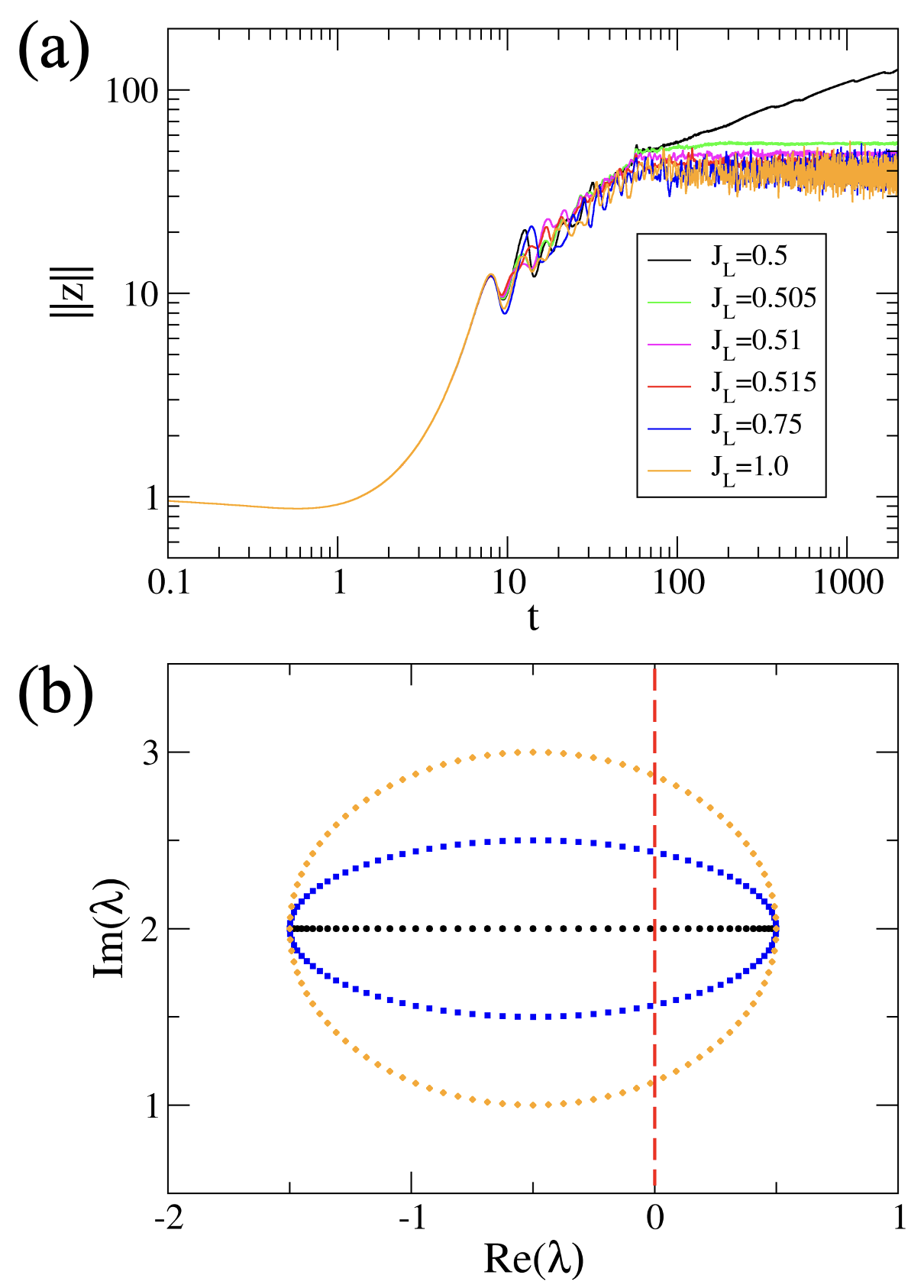}
    \caption{(a) Time evolution of the Euclidean norm and (b) complex eigenvalues of Jacobian matrix. The time evolution of the global Euclidean norm, Eq.~(\ref{eq:EucNorm}), corresponding to the spatiotemporal patterns in Fig~\ref{fig:fig3}. The curves illustrate a distinct pathway to spatiotemporal chaos in the dissipative--Kerr case. There is unbounded dynamics for symmetric case, while bounded fluctuation behaviors corresponding to spatiotemporal chaos increase as asymmetry increases. Parameters are the same as in Fig.~\ref{fig:fig1}.}
    \label{fig:fig4}
\end{figure}

\subsection{Dissipative coupling with Kerr-type nonlinearity}

We next analyze the case of dissipative coupling with Kerr-type nonlinearity, where the coupling constant is real ($K=1.0$) and the nonlinear coefficient is purely imaginary ($\xi = 0.1 i$). In contrast to the reactive case, dissipative coupling always broadens the distribution of the real parts of the Jacobian eigenvalues as $|K|$ increases. As a result, the maximal real part grows with $|K|$, independent of the sign of $K$, and once it becomes positive, the network is destabilized. This setting demonstrates how the loss of eigenfrequency coherence, triggered by coupling asymmetry, evolves into spatiotemporal chaos under the combined influence of dissipative coupling and Kerr-type nonlinearity.

Figure~\ref{fig:fig3} presents the corresponding dynamics. The initial condition is the same as in Fig.~\ref{fig:fig1}, with a single oscillator ($j=50$) initially excited. For the symmetric case ($J_L = J_R = 0.5$), the amplitudes of the oscillators increase over time because dissipative coupling generates unstable modes despite the negative linear growth rate $\mu=-0.5$. Due to the Kerr-type nonlinearity, the oscillators acquire different frequencies depending on initial conditions, even though the imaginary parts of the linear Jacobian eigenvalues are identical. 
This frequency dispersion induces amplitude differences among oscillators, and as a result, no convective wave pattern is formed. With small asymmetry, the spatiotemporal pattern develops weak irregularities, reflecting a gradual loss of eigenfrequency coherence related to traveling waves. As $J_L$ increases further, the pattern becomes strongly irregular, and the time traces of representative oscillators exhibit broadband fluctuations with chaotic dynamics. In this regime, nonlinear frequency shifts suppress divergence, reducing the effective growth rate through decoherence. In Fig.~\ref{fig:fig4}(a), the norm increases steadily when $J_L = J_R$. For the asymmetry, the norm increases and then approaches a plateau by the interplays between nonreciprocity and nonlinearity.

In contrast to the case of pure imaginary $K$, when the coupling constant $K$ is real,
corresponding to dissipative coupling, the eigenvalues of the Jacobian matrix reduce to
\begin{align}
\mathrm{Re}( \lambda_k ) &= \mu + K \cos q_k, \\
\mathrm{Im}( \lambda_k )&= \omega + K\,\Delta J \sin q_k .
\end{align}
In this case, the real parts lie within the finite interval $[\mu-|K|,\mu+|K|]$. 
For the symmetric choice $\Delta J=0$, all eigenvalues collapse onto a horizontal line at 
$\mathrm{Im}( \lambda_k ) = \omega$, indicating that the growth rates differ but the 
oscillation frequency remains fixed without nonlinearity [Fig.~\ref{fig:fig4}(b)]. 
Once asymmetry is introduced, the imaginary parts acquire a dependence on $q_k$, so that 
the straight line deforms into an ellipse oriented along the imaginary axis, 
characteristic of convective phenomena.

The Jacobian eigenvalue spectrum clarifies these observations. 
Since the maximal real part remains positive regardless of asymmetry, 
linear instability is always present. However, the interplay between nonreciprocal 
dissipative coupling and Kerr-type nonlinearity generates spatiotemporal chaos, 
prevents unbounded growth, and yields plateau-like saturation instead of divergence. 
We note that for the maximally asymmetric case (unidirectional coupling, $J_L=1.0$), 
the linear eigenvalue spectra reduce to circles in the complex plane centered at 
$(\mu,\omega)$. For finite $N$, dissipative and reactive couplings sample different 
angular positions on the circle, so their spectra appear distinct. In the large-$N$ 
limit, however, the spectra become identical in shape provided that the coupling 
strengths are equal (i.e., $|K|$ is the same), making the two cases indistinguishable 
by spectrum alone.

A further distinction arises in the role of nonreciprocity between the two Kerr-type cases. For reactive coupling, spatiotemporal chaos appears only after the asymmetry exceeds a finite threshold, marking a clear transition from convective decay to chaotic dynamics. In contrast, for dissipative coupling, increasing asymmetry immediately destroys eigenfrequency coherence, so that the system evolves into chaos without a threshold.

These analyses indicate that spatiotemporal chaos in our model arises only when three conditions are simultaneously satisfied: 
(i) the complex eigenvalues of the Jacobian matrix form an elliptic distribution in the complex plane, (ii) the maximal real part of these eigenvalues becomes positive, and (iii) the nonlinearity is of Kerr type. In the absence of any one of these conditions, the system does not develop fully chaotic dynamics but instead exhibits traveling waves, bounded periodic states, or unbounded divergence.

\begin{figure*}[!t]
    \centering
    \includegraphics[width=1.0\linewidth]{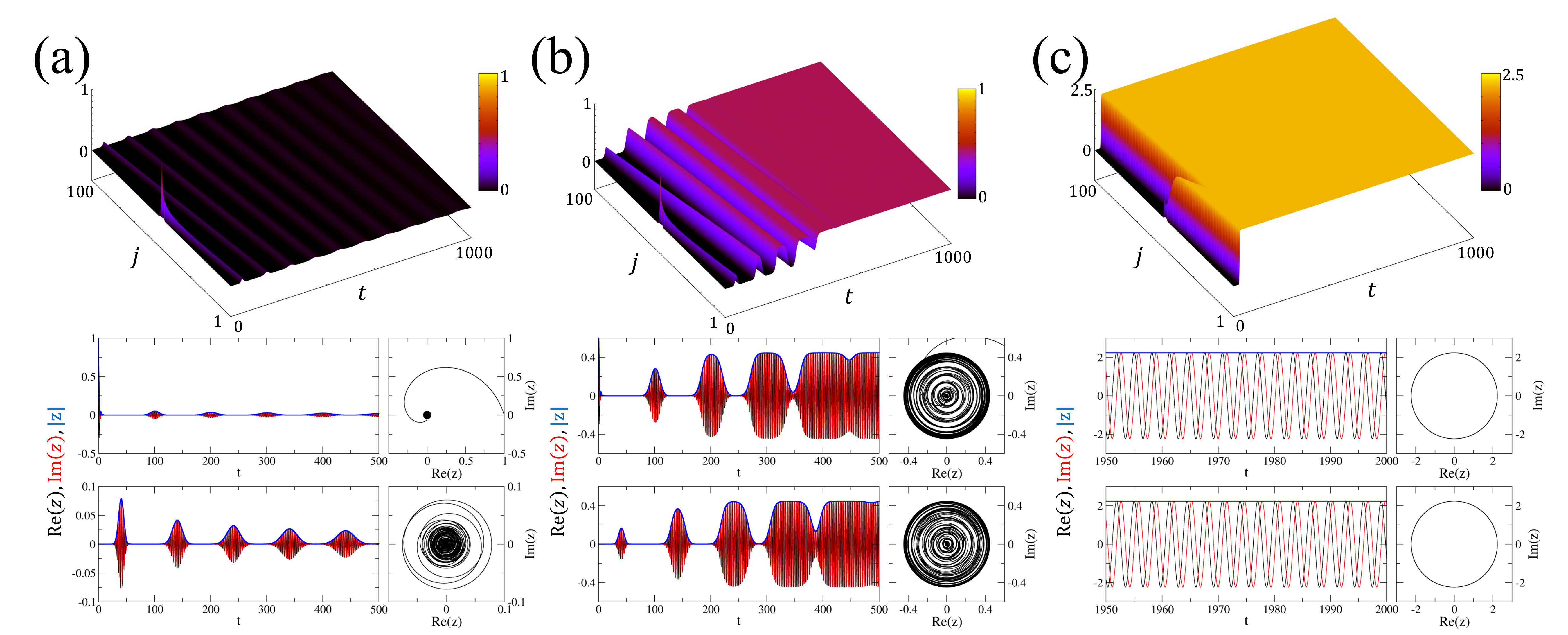}
    \caption{Reactive coupling with dissipative nonlinearity. Spatiotemporal amplitude patterns, oscillator time series, and trajectories in the complex plane for a ring of 100 Stuart--Landau oscillators. Parameters are $\mu=-0.5$, $\omega=2.0$, $\xi=0.1$, and $K=1.0i$. Panels show (a) $J_L=0.75$, (b) $J_L=0.76$, and (c) $J_L=1.0$. The amplitude plots show traveling waves that evolve into stable limit cycles. Here, nonlinear damping immediately balances growth, preventing chaos and enforcing bounded periodic oscillations.}
    \label{fig:fig5}
\end{figure*}

\begin{figure}[!t]
    \centering
    \includegraphics[width=0.5\linewidth]{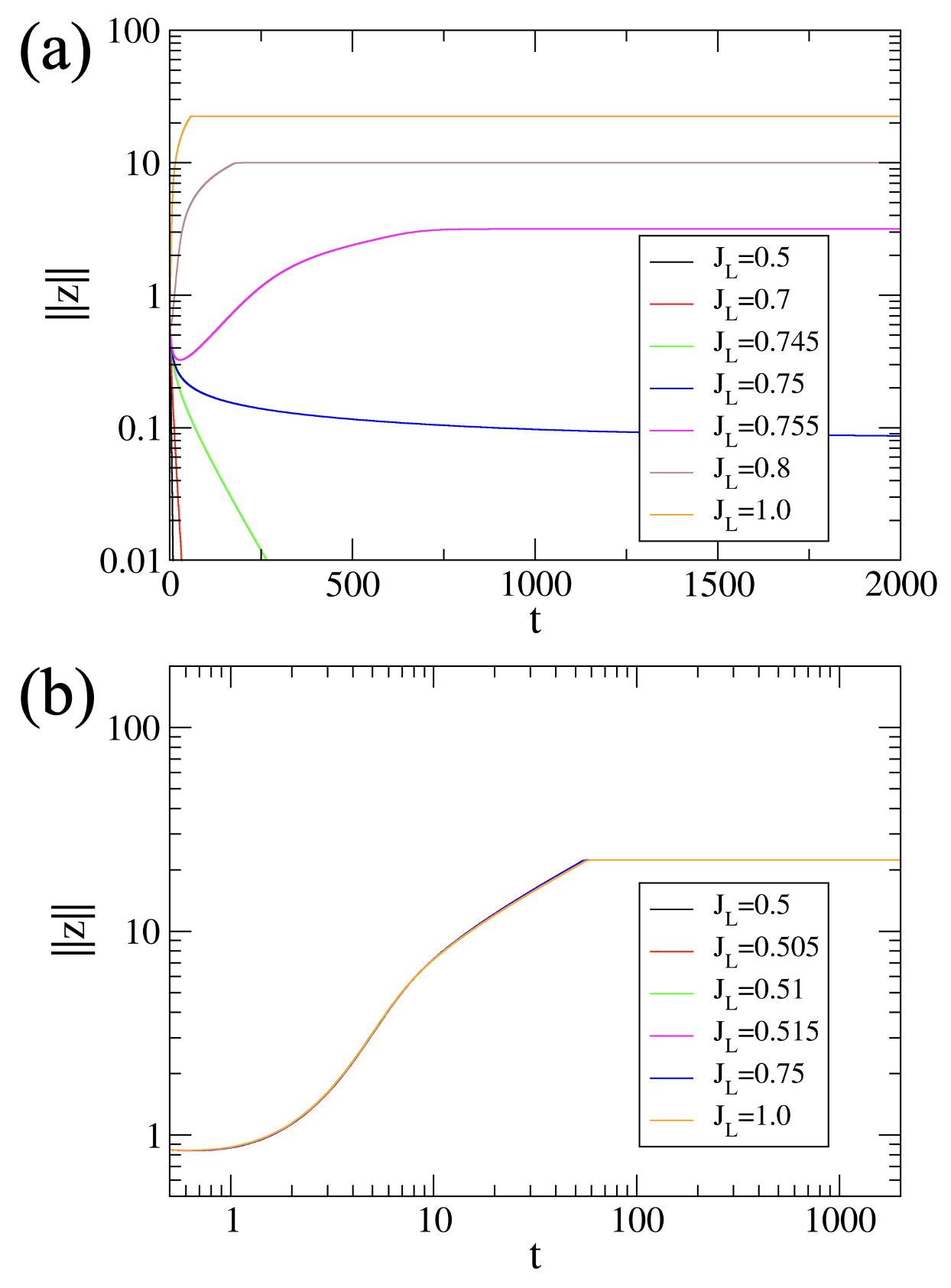}
    \caption{Time evolutions of the Euclidean norms. Time evolutions of the global Euclidean norm, Eq.~(\ref{eq:EucNorm}), correspond to the spatiotemporal patterns in Figs.~\ref{fig:fig5} and \ref{fig:fig7}. The curves illustrate distinct suppression mechanisms: (a) governed by nonreciprocity and (b) unaffected by it, respectively. Parameters are the same as in Fig.~\ref{fig:fig1}.
}
\label{fig:fig6}
\end{figure}

\subsection{Reactive coupling with dissipative nonlinearity}

We now consider reactive coupling with dissipative nonlinearity, 
where the coupling constant is purely imaginary ($K = 1.0 i$) and the nonlinear coefficient is real ($\xi = 0.1$). In this case the nonreciprocity generates convective instability and nonlinearity provides direct amplitude saturation, so they lead to limit cycles, unlike the case of Kerr-type nonlinearity.

Figure~\ref{fig:fig5} illustrates the resulting dynamics. For the symmetric case ($J_L = J_R = 0.5$), the amplitudes decay uniformly, and the system relaxes to a global amplitude death state due to the linear damping term. With small asymmetry, traveling waves appear in the spatiotemporal pattern and then they achieve global amplitude death through convective decay. When $J_L > 0.75$, traveling waves become unstable and evolve into limit cycles, whose amplitudes are immediately stabilized by nonlinear saturation. As a result, the dynamics consist of limit cycles rather than divergence and the amplitude increases as nonreciprocity increases. The time series of representative oscillators show limit cycles after transient, indicating stable finite-amplitude dynamics enforced by nonlinear damping. In Fig.~\ref{fig:fig6}(a), the norm remains bounded without fluctuation at all times, reflecting the immediate action of nonlinear amplitude saturation and the norm is systematically lower on average than that in Fig.~\ref{fig:fig2}(a), because the dissipative nonlinearity provides a stronger suppression of amplitude growth. The Jacobian eigenvalue spectrum exhibits the same deformation as in Fig.~\ref{fig:fig2}. The critical behaviors of instability are the same between them, dissipative nonlinearity does not achieve spatiotemporal chaos, contrary to the Kerr-type nonlinearity.

\begin{figure*}[!t]
    \centering
    \includegraphics[width=1.0\linewidth]{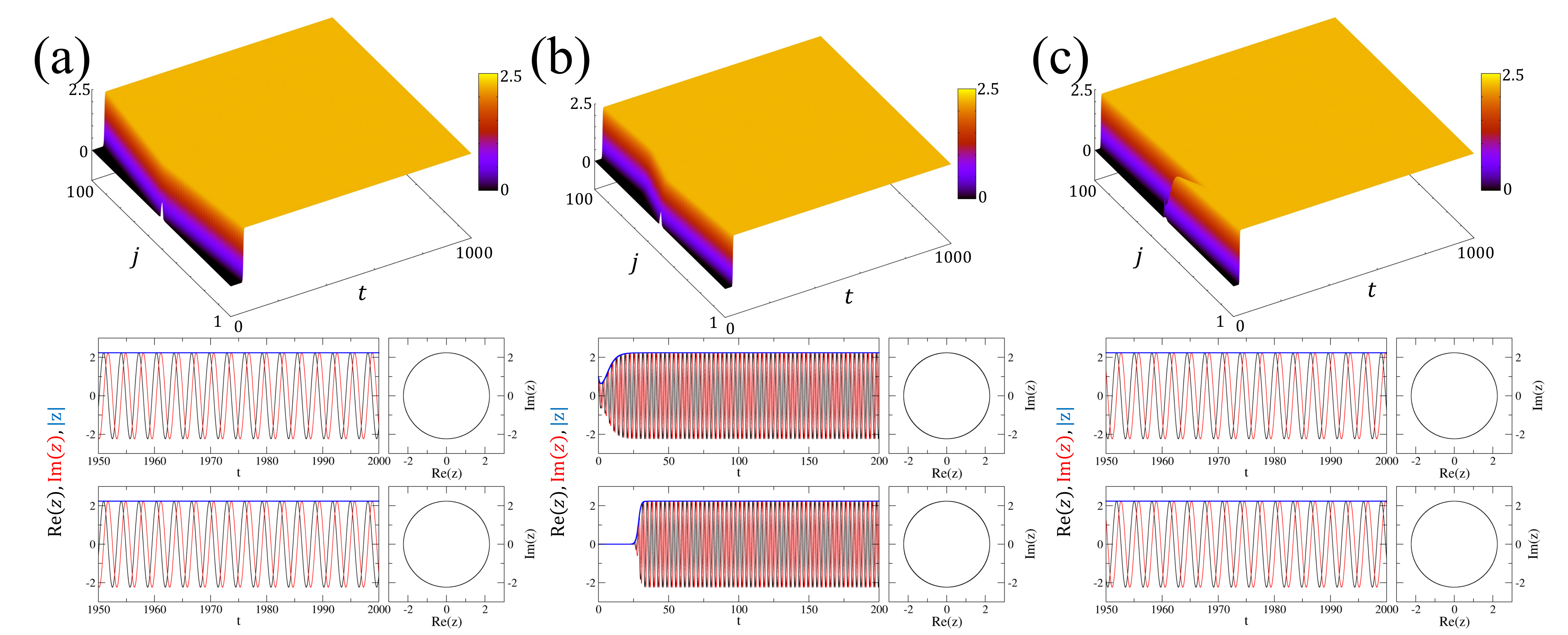}
    \caption{Dissipative coupling with dissipative nonlinearity. Spatiotemporal amplitude patterns, oscillator time series, and trajectories in the complex plane for a ring of 100 Stuart--Landau oscillators. Parameters are $\mu=-0.5$, $\omega=2.0$, $\xi=0.1$, and $K=1.0$. Panels show (a) $J_L=0.50$, (b) $J_L=0.75$, and (c) $J_L=1.0$. Despite the presence of unstable modes in the eigenvalue spectra, dissipative nonlinearity clamps amplitude growth, so the long-time dynamics remain finite and periodic limit cycles.}
    \label{fig:fig7}
\end{figure*}

\subsection{Dissipative coupling with dissipative nonlinearity}

Finally we analyze the case of dissipative coupling with dissipative nonlinearity, 
where both the coupling constant is real ($K=1.0$) and the nonlinear coefficient is real ($\xi = 0.1$). Here the nonlinearity provides direct amplitude saturation, 
while dissipative coupling promotes local amplitude growth. The competition between these effects produces bounded limit cycles rather than spatiotemporal chaos.

Figure~\ref{fig:fig7} summarizes the dynamics. For the symmetric case ($J_L = J_R = 0.5$), the oscillators grow in amplitude due to dissipative coupling, but the nonlinear saturation term limits the overall amplitude and stabilizes the system. As asymmetry is increased, the network always converges to limit cycles and the convergence to limit cycles is almost identical regardless of nonreciprocity. The Jacobian eigenvalue spectrum evolves analogously to Fig.~\ref{fig:fig4}. Although unstable modes with positive real parts of Jacobian matrix appear, the dissipative nonlinearity clamps amplitude growth, preventing divergence. Consequently, the long-time dynamics remain finite and periodic rather than chaotic. Finally, in Fig.~\ref{fig:fig6}(b), the norm also stays finite and converges to a stable asymptotic level, consistent with limit cycles after transient. In this case, enhanced nonreciprocity does not alter the dynamics, in contrast to the Kerr-type nonlinear case [Fig.~\ref{fig:fig4}(a)], where it enhances chaotic fluctuations. We note that for negative dissipative nonlinearity ($\mathrm{Re}(\xi)<0$), the nonlinear anti-damping destabilizes the system and the amplitude diverges.

\section{Discussion and Conclusion}

Our results demonstrate that nonreciprocity universally drives oscillator networks into spatiotemporal chaos when combined with Kerr-type nonlinearities, whereas dissipative nonlinearities suppress chaotic growth and enforce bounded periodic states. Nonreciprocal reactive coupling induces chaos through growth-rate modulation, whereas nonreciprocal dissipative coupling does so through coherence loss. Moreover, the reactive route exhibits a threshold-like transition, whereas the dissipative route drives the system into chaos without any threshold once asymmetry is introduced. Together these observations highlight our main message that nonreciprocity is not a minor perturbation but a fundamental organizing principle of complex dynamics. More precisely, chaos emerges only when the Jacobian eigenvalues form an elliptic distribution, the maximal real part becomes positive, and the system possesses Kerr-type nonlinearity.

This perspective has broader implications. It reveals how asymmetry reshapes stability landscapes across nonlinear systems and provides a minimal framework to anticipate when and how chaos emerges. Beyond oscillator networks, the same principles can be applied to active matter, photonic devices, and ecological or biological communities, where nonreciprocal interactions are ubiquitous. In such diverse contexts, the routes identified here can serve as predictive tools for either harnessing spatiotemporal chaos as a functional resource or suppressing it to maintain coherence and stability. Overall, our study establishes a compact theoretical foundation for understanding and controlling complexity in nonequilibrium systems driven by nonreciprocal interactions.




\section*{Acknowledgments}
The author acknowledges financial support from the Institute for Basic Science in the Republic of Korea through the project IBS-R024-D1.

\bibliography{nonreciprocal_refs}

\end{document}